\documentstyle[12pt,a4]{article}
\nopagebreak

\def\oneh{{\textstyle {1\over 2}}}
\def\treh{{\textstyle {3\over 2}}}
\def\pih{{\textstyle {\pi\over 2}}}

\begin{document}
\centerline{\Large\bf{Quantum interference effects in
($\vec{\mathrm e},{\mathrm e}'{\mathrm p}$) reactions}}
\vskip1.5cm
\centerline{\large{A.~Bianconi and S.~Boffi}}
\vskip0.5cm
\centerline{\small{Dipartimento di Fisica Nucleare e Teorica,
Universit\`a di Pavia, and}}

\centerline{\small{Istituto Nazionale di Fisica Nucleare,
Sezione di Pavia, Pavia, Italy}}

\vskip3.5cm


\begin{abstract}
The response to longitudinally polarized electrons in coincident
out-of-plane ($\vec{\mathrm e},{\mathrm e}'{\mathrm p}$) reaction
is discussed to study the role of final state interactions and
quantum interference between different reaction channels, i.e. between
direct (quasi) elastic proton emission and baryon resonance production.
The high-energy suppression of this effect due to colour transparency
is also discussed. Results are given for the
$^{12}$C($\vec{\mathrm e},{\mathrm e}'{\mathrm p}$)$^{11}$B$_{\mathrm
{g.s.}}$
reaction.
\end{abstract}
\clearpage


In quasifree nucleon knockout by longitudinally polarized electrons
(helicity $h=\pm 1$) the coincidence cross section is separated into
helicity independent ($\Sigma$) and dependent ($\Delta$) contributions
[1,2]. The part $\Sigma$ is responsible for nucleon
knockout by unpolarized electrons and has been extensively studied both
experimentally and theoretically (see references in [1]). The helicity
dependent part $\Delta$ is proportional to the socalled ``fifth'' structure
function, $W'_{\mathrm {LT}}$, and to $\sin\alpha$, where the
(out-of-plane) angle $\alpha$ is the angle of rotation about the momentum
${\vec q}$ of the exchanged virtual photon $\gamma^*$ to bring the electron
scattering plane to coincide with the plane of the final hadron system.
The fifth structure function $W'_{\mathrm {LT}}$ is the (imaginary part of the)
longitudinal-transverse (L-T) interference of the target currents and
vanishes identically in plane-wave-impulse approximation (PWIA) or
anyhow when final-state interactions (FSI) are absent. It also vanishes
when the reaction proceeds through a channel in which a single phase
dominates for all the projections of the target current [3]. As such,
$W'_{\mathrm {LT}}$ provides an observable which is highly sensitive to FSI.

The helicity dependent term $\Delta$ in the cross section breaks the
symmetry for particle emission above ($\alpha>0$) and below ($\alpha<0$)
the electron scattering plane for fixed electron helicity $h$ or,
alternatively, for electrons with $h=1$ and $h=-1$ and particles
emitted at a fixed $\alpha$. One may select $\alpha =\pi/2$ and
define the helicity asymmetry according to one of the
following alternative forms:

$$A={\Sigma\over\Delta}
= {N(1,\pih)- N(-1,\pih)\over N(1,\pih) + N(-1,\pih)}
= {N(1,\pih) - N(1,-\pih)\over N(1,\pih) + N(1,-\pih)},
$$
where $N(h,\alpha)$ is the number of events observed at a given angle
$\alpha$ for a fixed electron helicity $h$.

In a series of papers [4-7], the effect of the interference between
elastic and resonance channels has been studied in quasielastic scattering,
with special interest for the colour transparency problem [4,5,7] and for the
anomalous production of resonances in nuclei [6]. The idea is that in the
hard scattering with $\gamma^*$ the target nucleon, besides undergoing
elastic scattering, may also be excited into a barionic state N$^*$ [8].
Fermi motion plays an essential role in determining the composition of the
hadron wave packet [4,9]. In fact, the wave packet produced by absorption of
$\gamma^*$ is a superposition of physical hadronic states with the same
energy and direction of motion, but different mass and momentum magnitude.
These states propagate through the nuclear medium undergoing the soft FSI
which are then responsible for interference between the elastic and inelastic
channels. This modifies dramatically the cross section for particle emission.

In this paper, we investigate the effect of this kind of interference on the
helicity asymmetry $A$.

The time-independent coupled-channel formalism developed in [4,5] is here
extended to consider the case of three channels. The final-state wave
function is formed by three coupled components, labelled $i=1$, 2 and 3,
which, according to the analysis in [10], are taken to correspond to the
nucleon and two excited baryons, i.e. S$_{11}$(1535)
and F$_{15}$(1680), respectively.
These states are simultaneously produced with amplitudes $B_i$ in the hard
scattering at a given point ${\vec r}_0$. A common phase
$\exp({\mathrm i}{\vec q} \cdot {\vec r}_0)$ is given by the virtual photon
$\gamma^*$. Then we calculate a function of the form
$\exp({\mathrm i}{\vec q}\cdot{\vec r}_0)\cdot\Sigma_i B_i\Psi_i({\vec r},
{\vec r}_0)$ ($\Psi_i({\vec r}_0,{\vec r}_0)\equiv 1$), which is an
eigenfunction of the final state Hamiltonian $H_0 + V$ ($3\times 3$ matrix).
All the three channels describe particles propagating in the same direction
with the same energy, and $p_i = \sqrt{E^2 - M_i^2}$. $V\equiv \{V_{ij},
i,j=1, 2, 3\}$ is an optical potential describing FSI. Its nondiagonal terms
cause transitions between different channels. $V$ does not conserve flux
($V^\dagger \neq V$), but each of its elements conserves energy. The total
wave function  $\Psi({\vec r})$ is a coherent sum of all the waves of the
above kind emitted from any nuclear point ${\vec r}_0$.

The formalism in [5] is also modified to take into account the different
contributions of the longitudinal and transverse components of the target
current. The inital proton is assumed in the $j$-shell with total spin
component $m$ resulting from vector coupling of the intrinsic spin with
the orbital motion with angular momentum $l$.
Consequently, the transition matrix,
$M_{\mathrm {fi}}\ =\ M_{\mathrm {fi}}^{(+)} + M_{\mathrm {fi}}^{(-)}$,
is the sum of the two contributions corresponding to the two initial
(orbital) states $\vert l,m\mp\oneh\rangle$.

According to [5] the amplitudes $B_i(Q^2,s)$ are calculated as

$$\vert B_i(Q^2,s)\vert^2 = \int {\mathrm d}{\vec k}\,\vert g_i(Q^2)\vert^2
R_i(s)\,  n(k),$$
where $s \equiv (q+k)^\mu(q+k)_\mu$, $n(k)$ is the distribution for the
bound nucleon momentum,

$$R_i(s) = {{M_i\Gamma_i} \over {(s-M_i^2)+M_i^2\Gamma_i^2}}.
$$
The functions $g_i(Q^2)$ are the photocoupling amplitudes for the various
intermediate states produced by hard scattering. They were taken all equal
in [4-7] according to quark counting rules. In the present approach,
however, we are interested in the different spin transitions related to
longitudinal/transverse currents. Therefore $g_i(Q^2)$ must be taken
different for $M_{\mathrm {fi}}^{(\pm)}$ and expressed in terms of the helicity
amplitudes [11]:

$$g_i^{\pm}(Q^2)\equiv\langle i,m'_s\vert
{\vec e}'_h \cdot {\vec J}\vert  1;m_s = \pm 1/2\rangle,$$
where ${\vec J}$ is the target current and ${\vec e}'_h$ is the
virtual-photon polarization vector expressed in the reference frame with
the $z$-axis along the outgoing proton momentum ${\vec p}'$.

The photocoupling amplitudes are taken from the phenomenological helicity
amplitudes according to the analysis of [10,12].

The formalism has been applied to the reaction
$^{12}$C($\vec{\mathrm e},{\mathrm e}'{\mathrm p}$)$^{11}$B$_{\mathrm {g.s.}}$,
where the proton is
knocked out from the p-shell with $j=\treh$. This reaction has already been
studied under quasielastic conditions in the kinematic range available at
Bates with incident electrons of 560 MeV [13]. In this kinematic regime
conventional distorted-wave impulse approximation (DWIA) based upon
phenomenological optical potentials is quite successful in explaining the
cross section, the asymmetry and the extracted fifth structure function. The
sensitivity of $W'_{\mathrm {LT}}$ to FSI was proven in [13] to be useful in
disentangling different phase-shift equivalent optical potentials which
produce different scattering wave functions in the interior nucleus.

In this paper we are interested in quasielastic knockout ($\omega\sim
Q^2/2M$) in the region with $\vert{\vec q}\vert =$ 2--6 GeV. This region is
of interest for the planned experiments at CEBAF, and has already been
explored partially in the NE18 experiment at SLAC [14]. In the following,
results obtained from the full calculation with quantum interference will be
compared with PWIA and the conventional (Glauber) DWIA without quantum
interference.

Fig. 1 shows the ratio $R$ between the total number of events for
unpolarized incident electrons and the PWIA result at different values of the
longitudinal component $p_{m\parallel}$ (with respect to ${\vec q}$) of the
missing momentum ${\vec p}_m\equiv{\vec p}'-{\vec q}$. The transverse
component $p_{m\perp}$ is taken to be 200 MeV, a value comparable with the
Fermi momentum ($p_{\mathrm F} = 221$ MeV) and suitable to produce significant
out-of-plane yield. The figure confirms previously  obtained results [5,7],
where the role played by Fermi motion in producing the missing-momentum
dependence of the ratio $R$ has been emphasized. As already noticed in
[4,5], when one is not sensitive to ${\vec p}_m$ or effectively works at
$p_{m\parallel}\sim 0$, a flat ratio $R$ is obtained as a function of
$\vert{\vec q}\vert$ as in the NE18 experiment [14].

The main reason for calculating the ratio $R$ is to correlate it with the
results of the full coupled-channel calculation of $A$.
In the results for $R$, we can see three different
regimes; (1) no remixing at all, leading to curves that are PWIA times a
constant damping factor; (2) remixing without interference: at large ${\vec
p}_m\cdot \hat{\vec q}$ the traditional quasielastic yield is small, but many
$N^*$  are produced by the virtual photon, and converted into a proton  by
FSI; this leads to the prominent enhancement in $R$ that one can see in Fig.
1, and in [5,7]; (3) remixing with interference in the transition regions,
or at large $Q^2$. Actually, at large $Q^2$ and small ${\vec p}_m\cdot
\hat{\vec q}$ all the maxima for resonance production  overlap (see Fig. 3
in [5]), producing, hopefully, that ``pointlike'' hadron configuration that
should lead to colour transparency. At lower $Q^2$ remixing and interference
have been observed in a large enhancement of $N^*$ production in nuclei,
compared with $\Delta$ production in the same conditions [15,6].

In Fig. 2 the helicity asymmetry $A$ multiplied by $Q^2$ is plotted
for the conventional DWIA case and with quantum interference. All curves
become approximately constant as a function of $\vert \vec q\vert$. In fact
$W'_{\mathrm {LT}}$ contains combinations like $G_0 G_+$, where $G_\pm$ and
$G_0$
are the spin-flip and no-spin-flip helicity amplitudes for
$\gamma^* + {\mathrm p}\rightarrow {\mathrm p}$. At not too small $Q^2$ values,
$G_\pm/G_0\sim Q$, and in the total cross section the pure transverse
structure function $W_{\mathrm T}$ is the dominating term, so that $A$ becomes
proportional to $W'_{\mathrm {LT}}/W_{\mathrm T}$. The fact that $A$ is roughly
proportional to $1/Q^2$ suggests that in the considered range of $Q$ the
leading $Q$ dependence of the L-T interference cancels when taking the
appropriate combination to build $W'_{\mathrm {LT}}$.  Otherwise, one would
have $A\sim W'_{\mathrm {LT}}/W_{\mathrm T}\sim G_0 G_+ / G_+ G_+ \sim Q$.

In contrast, the quantum interference effects are evident. While $R$ is
especially sensitive to the regime (2) of remixing without interference,
the asymmetry $A$ is rather affected by regime (3).
Actually, a comparison of Figs. 1 and 2 shows that at small values of
${\vec p}_m \cdot \hat{\vec q}$ we obtain curves that present traditional
values of $R$, and for such  kinematics $A$ is not much affected by the
multichannel effects. On the contrary, at larger values of ${\vec p}_m
\cdot \hat{\vec q}$ the behaviour of $A$ is nontrivial. It is remarkable
that at large and positive ${\vec p}_m \cdot \hat{\vec q}$, but low $Q^2$,
the asymmetry can be enhanced by quantum interference with an excited
channel. In the high-$Q^2$ region, where onset of colour transparency could
start, the channels enter with relative weights which mutually cancel the
non-PWIA contributions, thus restoring the perfect PWIA, i.e. the perfect
nuclear transparency. In this regime no term dominates the asymmetry, which
goes to zero.

An interesting feature is the changing sign of $A$, which is clearly
correlated with the dominance of the inelastic channels as a consequence of
the choice of phases according to diffractive sum rules [4-7,17]. The
elastic and the inelastic channels enter the overall amplitude with
opposite signs. The dominance of the elastic or inelastic channels in
FSI directly manifests itself in the sign of the asymmetry. This occurs
at smaller values of $Q^2$ than in the ratio $R$, because  $A$ is
insensitive to the dominating PWIA-elastic contribution to the cross
section. This feature of $A$ is particularly remarkable, because a
direct measurement of transparency at the cross section level cannot
test the finite-energy sum rules nor determine the relative
{\sl phases}\/ of elastic and inelastic diffractive N-N scattering in FSI.

On the other side, the absolute values predicted here for $A$
are rather low. As a better guidance for a possible measurement,
in Fig. 3 the calculated helicity asymmetry $A$ is shown both in conventional
DWIA and with the quantum interference effects. A significant reduction is
introduced by such effects. However, the absolute magnitude of the asymmetry
is presumably too low to be observed. This low value can be explained with
the following argument. In conventional DWIA the fifth structure function
$W'_{\mathrm {LT}}$ is small because the longitudinal part is also small
and decreases with increasing ejectile energy faster than the transverse
part. Electromagnetic baryonic excitations are mostly transverse [12].
Thus the  L-T interference becomes even smaller when quantum interference
is taken into account. The same behaviour can be expected for the nucleon
recoil polarization $P_N$ in the direction normal to the hadron plane where
under parallel kinematic conditions (i.e. ${\vec p}'$ parallel to
${\vec q}$) a similar L-T structure function determines the observed
response (see also [17]).


\medskip

We are grateful to V.~Burkert and P.~Stoler for useful discussions.

\vfill
\clearpage


\vfill
\clearpage


\centerline{\large\bf Figure captions}

\medskip

Fig. 1. The transparency coefficient $R$ for the
$^{12}$C($\vec{\mathrm e},{\mathrm e}'{\mathrm p}$)$^{11}$B$_{\mathrm {g.s.}}$
reaction as a function of
the three-momentum transfer $q=\vert{\vec q}\vert$ (in GeV). The transverse
component $p_{m\perp}$ (with respect to ${\vec q}$) of the missing momentum
is fixed at 200 MeV. The longitudinal component is $p_{m\parallel}= 0, 100,
150, 200$ MeV for the solid, dotted, dashed and dot-dashed curves,
respectively.

\smallskip

Fig. 2. The helicity asymmetry $A$ multiplied by $Q^2$ for the
$^{12}$C($\vec{\mathrm e},{\mathrm e}'{\mathrm p}$)$^{11}$B$_{\mathrm {g.s.}}$
reaction as a function
of the momentum transfer  $q=\vert{\vec q}\vert$ (in GeV). Solid and dotted
curves calculated in conventional DWIA, dashed and dot-dashed curves
with quantum interference. Solid and dashed curves for a vanishing
longitudinal component $p_{m\parallel}$ of the missing momentum,
dotted and dot-dashed curves for $p_{m\parallel}= 200$ MeV. The
transverse component $p_{m\perp}$ (with respect to ${\vec q}$)
of missing momentum is fixed at 200 MeV.

\smallskip

Fig. 3. The helicity asymmetry $A$ for the
$^{12}$C($\vec{\mathrm e},{\mathrm e}'{\mathrm p}$)$^{11}$B$_{\mathrm {g.s.}}$
reaction as a function of three-momentum
transfer $q=\vert{\vec q}\vert$ (GeV) for a missing momentum with 200 MeV
transverse and 100 MeV longitudinal components with respect to the photon
momentum ${\vec q}$.  Solid and dashed (dotted and dot-dashed) curves for an
electron scattering angle $\theta=36^{\mathrm o}$ ($72^{\mathrm o}$). Dashed
and
dot-dashed curves in conventional DWIA, solid and dotted curves with quantum
interference of  intermediate hadron states.


\end{document}